\documentclass[12pt]{article}
\usepackage{epsfig,amssymb,amsmath,psfrag,hyperref}
\usepackage[utf8]{inputenc}
\usepackage[english]{babel}
\usepackage{color}
\hypersetup{
  linktocpage,
  colorlinks  = true, 
  urlcolor    = blue, 
  linkcolor   = blue, 
  citecolor   = red 
}

\def \be  {\begin{equation}}
\def \ee  {\end{equation}}
\def \ba  {\begin{eqnarray}}
\def \ea  {\end{eqnarray}}

\def \o{\mathcal{O}}
\newcommand \widebar [1] {\overline{#1}}
\def\x{x}
\def\xb{\bar{x}}
\def\dbar#1{\widebar{D}_{#1}}
\def\me{\mathcal{M}}

\def\l{\lambda}
\def\ll{\lambda^{-\frac{3}{2}}}
\def\lll{\lambda^{-\frac{5}{2}}}

\def \ot{\mathcal{O}_2}
\def \ott{\mathcal{O}_3}
\def \op{\mathcal{O}_p}
\def \corr22pp{\langle\ot\ot\op\op\rangle}
\def \fourtwo{\langle\ot\ot\ot\ot\rangle}

\def \cpi{C_{pp,i}}

\def \ca{\mathcal{A}}
\begin{document}
\thispagestyle{empty}

\null\vskip-12pt \hfill  \\
\null\vskip-12pt \hfill   \\

\vskip2.2truecm
\begin{center}
\vskip 0.2truecm {\Large\bf
{\Large One-loop string corrections for AdS Kaluza-Klein amplitudes}
}\\
\vskip 1truecm
{\bf J.~M. Drummond, R. Glew and H.~Paul\\
}

\vskip 0.4truecm
 
{\it
 School of Physics and Astronomy, University of Southampton,\\
 Highfield, Southampton SO17 1BJ\\
\vskip .2truecm                        }
\end{center}

\vskip 1truecm 
\centerline{\bf Abstract}\normalsize
We discuss the string corrections to one-loop amplitudes in AdS$_5\times$S$^5$, focussing on their expressions in Mellin space. We present the leading $(\alpha')^3$ corrections to the family of correlators $\langle \mathcal{O}_2 \mathcal{O}_2 \mathcal{O}_p \mathcal{O}_p \rangle$ at one loop and begin the exploration of the form of correlators with multiple channels. From these correlators we extract some string corrections to one-loop anomalous dimensions of families of operators of low twist.

\medskip
\noindent                                   
\newpage
\setcounter{page}{1}\setcounter{footnote}{0}
\tableofcontents
\section{Introduction}\setcounter{equation}{0}
Recently, significant progress has been made in understanding the structure of scattering amplitudes in anti-de-Sitter space by analysing the dual conformal field theory. Many studies in this direction have been made in a number of papers, analysing both tree and loop diagrams in AdS. A number of techniques have been employed but the most important tool in all these developments has been to impose consistency of the operator product expansion of the boundary CFT.

Many investigations have focussed on the example of type IIB string theory on AdS$_5\times$S$^5$, dual to $\mathcal{N}=4$ super Yang-Mills theory. The interesting physical quantities of this theory can be studied in a double expansion in $1/N$ and $\lambda^{-\frac{1}{2}}$ around the supergravity limit, $N \gg \lambda \gg 0$. The leading terms in the $1/N$ expansion come from disconnected free field diagrams and are independent of $\lambda$. The first $1/N^2$ corrections arise from tree-level string interactions. At leading order in $\lambda^{-\frac{1}{2}}$ these were given in \cite{Rastelli:2016nze,Rastelli:2017udc} and take a particularly simple form in a Mellin representation. These results allow a resolution of the mixing of the spectrum of degenerate double trace operators which control the operator product expansion at this order, yielding a simple formula for the leading contributions to their anomalous dimensions. Recently there has been significant progress in understanding the nature of the $\lambda^{-\frac{1}{2}}$ corrections at tree level \cite{Alday:2018pdi,Binder:2019jwn,Drummond:2019odu,Drummond:2020dwr}. Here it is very interesting to note that the anomalous dimensions of double-trace operators receive a very restricted set of $\lambda^{-\frac{1}{2}}$ corrections~\cite{Drummond:2019odu,Drummond:2020dwr}.

The next corrections in the large $N$ expansion correspond to one-loop amplitudes. The supergravity contributions have been investigated from the CFT perspective in \cite{Alday:2017xua,Aprile:2017bgs_2222,Alday:2017vkk,Aprile:2017qoy_2233,Aprile:2019rep,Alday:2019nin} while string corrections have also been investigated in \cite{Alday:2018pdi,Alday:2018kkw,Drummond:2019hel}. So far the Mellin space and position space representation for the $\lambda^{-\frac{3}{2}}$ corrections to the one-loop amplitudes is known for the simplest correlator, that of four stress-energy multiplets, dual to four-graviton scattering in AdS${}_5$.

The aim of this analysis is to calculate more general one-loop correlators of four 1/2-BPS operators at order $\lambda^{-3/2}$. Here we consider the correlator of two stress-energy multiplets and two Kaluza-Klein modes denoted here by $\langle \mathcal{O}_2 \mathcal{O}_2 \mathcal{O}_p \mathcal{O}_p \rangle^{(2,3)}$ which gives an infinite family of correlators with only a single $su(4)$ channel. We also consider the first example of a correlator with multiple $su(4)$ channels $\langle \mathcal{O}_3 \mathcal{O}_3 \mathcal{O}_3 \mathcal{O}_3 \rangle$.  The information needed for these one-loop calculations is already encoded in known tree-level data. In particular, due to the very special structure of the $\lambda^{-\frac{1}{2}}$ corrections to the double-trace spectrum, the leading double logarithmic discontinuity at one loop can be obtained through the action of a simple differential operator on the discontinuity of the corresponding tree-level amplitude. With this to hand, it is simple to construct the basis of the Mellin representation. In contrast to the supergravity results presented in \cite{Alday:2019nin}, it is then necessary to fix an additional set of simple pole contributions which turn out to have a very simple form.

The rest of this paper is organised as follows. We start with a general discussion of the $\corr22pp$ four-point correlation function in the supergravity limit, and discuss their structure both in position space as well as in their Mellin space representation. In Section~\ref{sec:one_loop_Mellin}, we specialise to one-loop Mellin amplitudes. We first review the known one-loop results for supergravity amplitudes, and then present how to generalise the one-loop string corrected Mellin amplitude at order $\ll$ to the $\corr22pp$ family of correlators, which is the main result of this work. We will explain that the double discontinuity is not enough to fully fix the Mellin amplitude, and we need to add an à priori unbounded tower of additional window poles (which were shown to be absent in the supergravity case). In Section~\ref{sec:window_poles}, we demonstrate how these window poles are fixed by tree-level data. Interestingly, it turns out that at most only five extra poles are necessary and they follow a simple pattern. In Section~\ref{sec:3333} we explicitly construct the one-loop $\ll$ correction to the $\langle\ott\ott\ott\ott\rangle$ correlator, which is the first correlator with non-trivial $(\sigma,\tau)$ dependence. In Section~\ref{sec:twist5_anom_dim} we extract new subleading CFT data from our results. Finally, in Section~\ref{sec:flat_space_limit} we consider the flat space limit and show that our new results correctly match the non-analytic part of the ten-dimensional one-loop string-amplitude.

\subsection{The $\corr22pp$ correlator}
We will consider four-point correlation functions of protected one-half BPS operators, which according to the AdS/CFT correspondence describe scattering amplitudes in AdS$_5\times$S$^5$. The operators dual to single-particle states in AdS are not simply one-half BPS single-trace operators but they require admixtures of multi-trace operators which are $1/N$ suppressed:\footnote{This subtlety was already noticed in the early works~\cite{DHoker:1999jke,Arutyunov:2000ima} and discussed more recently again in~\cite{Rastelli:2017udc,Aprile:2018efk,Arutyunov:2018neq,Aprile:2019rep,Alday:2019nin,Aprile:2020uxk}. } 
\begin{align}\label{eq:single_particle_op}
	\op = y^{i_1}\cdots y^{i_p}\text{Tr}\big(\Phi_{i_1}\cdots\Phi_{i_p}\big) + \text{(multi-traces)},
\end{align}
where $\Phi_i$ are the scalar fields of the $\mathcal{N}=4$ multiplet and $y^i$ are auxiliary $so(6)$ vectors obeying the null condition $y\cdot y=0$, such that $\op$ transforms in the traceless symmetric representation $[0,p,0]$ and has protected scaling dimension $\Delta=p$. In the holographic context, $\ot$ is dual to the scalar in the graviton supermultiplet, whereas the single-particle operators $\op$ with $p\geq3$ are dual to supergravity Kaluza-Klein modes which arise from compactification on S$^5$.

As discussed first in~\cite{Aprile:2018efk}, the multi-trace terms in the definition of the supergravity single-particle operators~\eqref{eq:single_particle_op} are such that \textit{single-particle operators $\op$ are orthogonal to all multi-trace operators}, i.e. $\langle\op\left[\o_{q_1}\ldots\o_{q_n}\right]\rangle=0$. This definition via orthogonality of operators allows one to compute the additional multi-trace terms purely within free field theory, and the results are \textit{exact in $N$}. For example, the first single-particle operator with a multi-trace admixture is the dimension four operator $\o_4$ given by~\cite{Aprile:2018efk}
\begin{align}
	\o_4 = \text{Tr}\big(\Phi^4\big)-\frac{2N^2-3}{N(N^2+1)}\ot\ot \,.
\end{align}
A general formula for the multi-trace terms of all single-particle operators $\op$ has been recently given in~\cite{Aprile:2020uxk}. The two-point functions of single-particle operators take the form
\be
\langle \mathcal{O}_p(x_1,y_1) \mathcal{O}_p(x_2,y_2) \rangle = g_{12}^p R_p(N)
\ee
where $g_{ij} = (y_{ij}^2/x_{ij}^2) $ is the propagator and $y_{ij}^2 \equiv y_i \cdot y_j$. Note that since the operators $\mathcal{O}_p$ are half-BPS, $R_p$ is independent of the coupling $g_{\rm YM}$ or $\lambda$.
The operators are normalised so that in the large $N$ limit we have $R_p \rightarrow p N^p$, e.g. $R_2 = 2(N^2-1)$. Many more properties of single-particle operators are described in~\cite{Aprile:2020uxk}.

To discuss four-point functions it is helpful to introduce the conformal and $su(4)$ R-symmetry cross-ratios,
\begin{align}
	u= x \xb &= \frac{x_{12}^2x_{34}^2}{x_{13}^2 x_{24}^2}, &&v=(1-x)(1-\xb)=  \frac{x_{14}^2x_{23}^2}{x_{13}^2 x_{24}^2}, \notag \\
	\frac{1}{\sigma}=y \bar y &= \frac{y_{12}^2 y_{34}^2}{y_{13}^2 y_{24}^2}, &&\frac{\tau}{\sigma}=(1-y)(1-\bar y)=\frac{y_{14}^2 y_{23}^2}{y_{13}^2 y_{24}^2}.
	\label{eq:crossratios}
\end{align}
In this paper, we will mainly focus on the four-point correlation function of two stress-tensor superprimary operators $\ot$ and two Kaluza-Klein modes $\op$. Superconformal symmetry constrains these correlators to take the form~\cite{Eden:2000bk,Nirschl:2004pa}
\begin{align}\label{eq:superconformal_constraint}
	\corr22pp = \corr22pp_{\text{free}} + \frac{1}{2p} R_2 R_p g_{12}^2g_{34}^p~\mathcal{I}~\mathcal{H}_p(u,v),
\end{align}
where the combination $g_{12}^2g_{34}^p$ carries the correct conformal weight and the $so(6)$ R-symmetry weights $y_i$ of the correlator. The free theory correlator takes the form
\be
\langle \mathcal{O}_2 \mathcal{O}_2 \mathcal{O}_p \mathcal{O}_p \rangle_{\rm free} = R_2 R_p g_{12}^2 g_{34}^p  \biggl[1 + \delta_{2p}\Bigl[u^2\sigma^2 +  \frac{u^2 \tau^2}{v^2}\Bigr] + 2 p a \Bigl[u\sigma + \frac{u\tau}{v} +(p-1)\frac{u^2 \sigma \tau}{v}\Bigr]\biggr]\,.
\label{freetheory}
\ee
Here we introduced $a = 1/(N^2-1)$ which we will take as our large $N$ expansion parameter. It is particularly convenient to do so as the free theory contribution then has exactly two terms in this expansion upon factoring out the normalisation $R_2 R_p$. Being the free theory contribution, no term in (\ref{freetheory}) depends on $\lambda$.

The factor $\mathcal{I}$ in the second term in (\ref{eq:superconformal_constraint}) is fixed by superconformal Ward identities and takes the factorised form
\begin{align}
	\mathcal{I} = \frac{(x-y)(x-\bar{y})(\xb-y)(\xb-\bar{y})}{(y\bar{y})^2}\,.
	\label{calI}
\end{align}
Finally, the factor $\mathcal{H}_p(u,v)$ is the only part of the correlator which depends on the gauge coupling $g_{\text{YM}}$ (or $\lambda$) and we will therefore refer to it as the interacting part of the correlator. It contains all the non-trivial dynamical information of the theory, and for this reason it receives contributions from unprotected operators only. For the special case of the $\corr22pp$ family of correlators, $\mathcal{H}_p(u,v)$ is independent of the internal $so(6)$ variables $(\sigma,\tau)$ and only a function of the conformal cross-ratios. It obeys the crossing transformations
\begin{align}\label{eq:crossing_trafo}
	\mathcal{H}_p(u,v)=\frac{1}{v^2}\mathcal{H}_p(u/v,1/v),\qquad\mathcal{H}_2(u,v)=\frac{u^2}{v^2}\mathcal{H}_2(v,u),
\end{align}
where the second relation is due to the enhanced crossing symmetry of the $\fourtwo$ correlator.

We will consider the expansion of $\mathcal{H}_p(u,v)$ around the supergravity limit, where one first takes the large $N$ limit (keeping the 't Hooft coupling $\l=g_{\text{YM}}^2N$ fixed) and then expands around large $\l$. In this limit, the interacting part admits the double expansion
\begin{align}\label{eq:double_expansion}
	\mathcal{H}_p &=  ~a \bigl( \mathcal{H}_p^{(1,0)}+\ll\mathcal{H}_p^{(1,3)}+\lll\mathcal{H}_p^{(1,5)}+ \lambda^{-3} \mathcal{H}_p^{(1,6)} + \ldots\bigr) 
	\notag \\
	& + a^2 \bigl( \lambda^{\frac{1}{2}} \mathcal{H}_p^{(2,-1)} + \mathcal{H}_p^{(2,0)} + \lambda^{-\frac{1}{2}} \mathcal{H}_p^{(2,1)} +\lambda^{-1} \mathcal{H}_p^{(2,2)}+\ll\mathcal{H}_p^{(2,3)}+\ldots\bigr)+O(a^3).
\end{align}

The term of order $a^0$ in (\ref{freetheory}) is the contribution from disconnected free field theory. At order $a$, we have the contribution from connected free field theory as well as the contribution from the interacting part given by the first line of (\ref{eq:double_expansion}). The term $\mathcal{H}_p^{(1,0)}$ and the order $a$ contribution from free field theory together correspond to the contribution of tree-level supergravity. These are then followed by an infinite tower of $\lambda^{-\frac{1}{2}}$ corrections $\mathcal{H}_p^{(1,n)}$ which arise from contact interaction vertices in the string theory effective action. These tree-level terms are most conveniently studied in their Mellin space representation which we introduce in Section~\ref{sec:Mellin_space_representation}. They are currently known up to and including the order $\lambda^{-\frac{5}{2}}$ terms \cite{Alday:2018pdi,Binder:2019jwn,Drummond:2019odu,Drummond:2020dwr}.

The order $a^2$ terms of the double expansion (\ref{eq:double_expansion}) correspond to one-loop amplitudes in AdS$_5$. Note that the leading term $\lambda^{\frac{1}{2}} \mathcal{H}_p^{(2,-1)}$ corresponds to the presence of a quadratic divergence at one loop in ten-dimensional supergravity. This divergence is regulated by a specific $\mathcal{R}^4$ counterterm at one loop in string theory. The term $\mathcal{H}_p^{(2,0)}$ is the one-loop supergravity term, as addressed in \cite{Aprile:2017bgs_2222,Alday:2017vkk,Aprile:2017qoy_2233,Aprile:2019rep,Alday:2019nin}. The term $\lambda^{-\frac{1}{2}} \mathcal{H}_p^{(2,1)}$ corresponds to the genus-one contribution to the modular completion of the $\lambda^{-\frac{5}{2}} \mathcal{H}_p^{(1,5)}$ term. The corresponding modular function is an Eisenstein series which receives perturbative contributions only at genus zero and genus two \cite{Green:1999pu} and we therefore expect $\mathcal{H}_p^{(2,1)}$ to vanish. The vanishing of this term is also consistent with the localisation analysis of \cite{Chester:2019pvm,Chester:2020dja}. The term $\mathcal{H}_p^{(2,2)}$ gives rise, in the flat space limit, to the analytic part of the one-loop string amplitude studied in \cite{Green:1999pv}. It is therefore non-vanishing and it corresponds to the genus-one contribution to the modular completion of the $\lambda^{-3} \mathcal{H}_p^{(1,6)}$ term. The next term $\lambda^{-\frac{3}{2}} \mathcal{H}_p^{(2,3)}$ is the genuine one-loop string correction induced by the presence of the $\lambda^{-\frac{3}{2}} \mathcal{H}_p^{(1,3)}$ term at tree level. This term is the one which we will construct in this paper.

The position space structure of one-loop IIB supergravity amplitudes has been addressed in~\cite{Alday:2017xua,Aprile:2017bgs_2222,Aprile:2017qoy_2233}, culminating in a general algorithm for constructing correlators with arbitrary external charges~\cite{Aprile:2019rep}. Considering further string corrections at one-loop has revealed a new type of singularity in their analytic structure compared to the supergravity case~\cite{Drummond:2019hel}. A complementary approach to one-loop amplitudes using their Mellin space representation is reviewed in Section~\ref{sec:one_loop_Mellin}.\\

Let us now describe the superconformal block decomposition of the unprotected part of a correlator. Note that in some intermediate calculations presented in Section~\ref{sec:window_poles} we will also make use of the block decomposition of more general correlators of the form $\langle\o_p\o_p\o_q\o_q\rangle$. After projecting onto unprotected singlets, the superconformal block decomposition reads
\begin{align}\label{eq:block_deco}
	\langle\o_p\o_p\o_q\o_q\rangle|^{\text{long}}_{[0,0,0]} = g_{12}^pg_{34}^q~\mathcal{I}~\sum_{t,\ell} C_{pp;t,\ell} C_{qq;t,\ell} \, G_{t,\ell}(\x,\xb),
\end{align}
where the sum runs over all exchanged unprotected primary operators with half-twist $t\equiv(\Delta-\ell)/2$ and even spin $\ell$. The functions $G_{t,\ell}(\x,\xb)$ are simply related to conformal blocks and are fixed by conformal symmetry to take the form~\cite{Dolan:2000ut,Dolan:2003hv}
\begin{align}\label{eq:conformal_block}
	G_{t,\ell}(\x,\xb) = (-1)^\ell (\x\xb)^t~\frac{\x^{\ell+1}F_{t+\ell+2}(\x)F_{t+1}(\xb)-\xb^{\ell+1}F_{t+\ell+2}(\xb)F_{t+1}(\x)}{\x-\xb},
\end{align}
where $F_{\rho}(\x)={}_2F_1\left(\rho,\rho,2\rho;\x\right)$ is the standard hypergeometric function.

The parameters $a$ and $\lambda^{-\frac{1}{2}}$ enter through the quantities $C_{pp}$ and the dimensions $\Delta$. In the supergravity limit and to leading order in $a$, the spectrum of exchanged operators is given by a set of unprotected double-trace operators with classical dimension $\Delta^{(0)}=2t+\ell$ and spin $\ell$. It turns out that generically there are many such operators with the same classical quantum numbers, which leads to a mixing problem. The quantum numbers $(t,\ell)$ are thus insufficient to fully describe the set of exchanged double-trace operators, and we are led to introduce an additional degeneracy label $i$, where in the singlet channel $i=1,\ldots,t-1$.\footnote{In the singlet channel, the set of $t-1$ degenerate double-trace operators is of the schematic form
\begin{align*}
 \left\{\o_2\square^{t-2}\partial^\ell\o_2,~\o_3\square^{t-3}\partial^\ell\o_3,~\ldots~,~\o_t\square^0\partial^\ell\o_t\right\}.
\end{align*}} 
We denote the (canonically normalised) scaling eigenstates by $\mathcal{K}_i$. The $C_{pp,i}$ are then related to the three-point functions $\langle \mathcal{O}_p \mathcal{O}_p \mathcal{K}_i \rangle$ and admit the double expansion
\begin{align}\label{eq:c_expansions}
	\cpi =	~~~&\left(\cpi^{(0,0)} + \ll \cpi^{(0,3)} + \ldots\right) \notag \\
		 	 +a&\left( \cpi^{(1,0)} + \ll \cpi^{(1,3)} + \ldots\right) + O(a^2).
\end{align}
Similarly their scaling dimensions admit the expansion,
\begin{align}\label{eq:anom_dims_expansion}
	\Delta_i = \Delta^{(0)} + 2a&\left(\eta^{(1,0)}_i + \ll\eta^{(1,3)}_i + \lll\eta^{(1,5)}_i + \ldots\right)\notag \\
			+2a^2&\left(\lambda^{\frac{1}{2}} \eta_i^{(2,-1)} +  \eta^{(2,0)}_i + \lambda^{-\frac{1}{2}} \eta^{(2,1)}_i + \lambda^{-1} \eta^{(2,2)}_i + \ll\eta^{(2,3)}_i + \ldots\right)+O(a^3).
\end{align}
At the order of tree-level supergravity the mixing problem described above has been solved in~\cite{Aprile:2017xsp_unmixing,Aprile:2018efk}, where a surprisingly simple and fully factorised formula for the supergravity anomalous dimensions $\eta^{(1,0)}_i$ was found. Also the first string correction exhibits a simple pattern in its anomalous dimensions $\eta^{(1,3)}_i$: only the lightest state (with degeneracy label $i=1$) acquires a non-vanishing anomalous dimension, and furthermore the first string correction to the three-point functions $\cpi$ was shown to vanish~\cite{Drummond:2019odu}:
\begin{align}
	\cpi^{(0,3)} = 0.
\end{align}
We have also included terms in the anomalous dimensions proportional to positive powers of $\lambda^{\frac{1}{2}}$. These account for contributions to $\mathcal{H}_p^{(2,-1)}$ at one loop.

In Section~\ref{sec:window_poles}, we will derive some new results for the unmixed three-point functions $\cpi^{(1,0)}$ and $\cpi^{(1,3)}$. Finally, in Section~\ref{sec:twist5_anom_dim} we compute the one-loop string corrected anomalous dimensions $\eta_i^{(2,3)}$ for some low twist operators.

\subsection{The Mellin space representation}\label{sec:Mellin_space_representation}
In the context of holographic correlators, it was found that comparably simple structures emerge when considering the correlator in its Mellin space representation. For the $\corr22pp$ family of correlators, the (reduced) Mellin amplitude $\me_p(s,t)$ of the interacting part $\mathcal{H}_p(u,v)$ is defined through the integral transform
\begin{align}\label{eq:Mellin_definition}
	\mathcal{H}_p(u,v)=\int_{-i\infty}^{i\infty}\frac{ds}{2}\frac{dt}{2}u^{\frac{s}{2}}v^{\frac{t-2-p}{2}}\me_p(s,t)\Gamma\big[\tfrac{4-s}{2}\big]\Gamma\big[\tfrac{2p-s}{2}\big]\Gamma\big[\tfrac{2+p-t}{2}\big]^2\Gamma\big[\tfrac{2+p-\tilde{u}}{2}\big]^2,
\end{align}
where the Mellin variables $(s,t,\tilde{u})$ obey $s+t+\tilde{u}=2p$. The crossing transformations~\eqref{eq:crossing_trafo} of $\mathcal{H}_p(u,v)$ translate directly into symmetries of the corresponding Mellin amplitudes:
\begin{align}\label{eq:crossing_trafo_Mellin}
	\me_p(s,t)=\me_p(s,\tilde{u}),\qquad\me_2(s,t)=\me_2(t,s).
\end{align}
Furthermore, the Mellin amplitudes $\me_p(s,t)$ admit an analogous double expansion as given in~\eqref{eq:double_expansion} for the correlator in position space:
\begin{align}\label{eq:double_expansion_Mellin}
	\me_p &= ~a\bigl( \me_p^{(1,0)}+\ll\me_p^{(1,3)}+\lll\me_p^{(1,5)}+\lambda^{-3}\me_p^{(1,6)}\ldots\bigr)
		\notag \\
	   	& + a^2\bigl(\lambda^{\frac{1}{2}} \me_p^{(2,-1)} + \me_p^{(2,0)}+ \lambda^{-\frac{1}{2}} \me_p^{(2,1)}+ \lambda^{-1} \me_p^{(2,2)} +\ll\me_p^{(2,3)}+\ldots\bigr)+O(a^3).
\end{align}

At tree-level, both the supergravity amplitude $\me^{(1,0)}$~\cite{Rastelli:2016nze,Rastelli:2017udc} and its first string correction $\me^{(1,3)}$~\cite{Drummond:2019odu} are known for correlators with arbitrary external charges. At the next order in $1/\lambda$, progress was made using various methods: in~\cite{Alday:2018pdi} the flat space limit was used to constrain the amplitude $\me_p^{(1,5)}$, whose final form was later fixed using supersymmetric localisation~\cite{Binder:2019jwn} and re-derived in~\cite{Drummond:2019odu} by considering constraints on the spectrum of anomalous dimensions $\eta^{(1,5)}_i$. Recently, the generalisation to correlators of arbitrary charges was achieved using bootstrap methods in Mellin space~\cite{Drummond:2020dwr}. The structure of one-loop Mellin amplitudes, and in particular the first one-loop string correction $\me_p^{(2,3)}$ which is the main focus of this paper, will be reviewed in the next section.

Lastly, note that in the limit of large $s$ and $t$ the Mellin amplitude is related to physics in ten-dimensional flat space as we discuss in Section~\ref{sec:flat_space_limit}.
\section{One-loop Mellin amplitudes}\setcounter{equation}{0}\label{sec:one_loop_Mellin}
\subsection{One-loop supergravity}\label{sec:one_loop_sugra}
A complimentary approach to the position space approach developed in~\cite{Aprile:2017bgs_2222,Aprile:2017qoy_2233,Aprile:2019rep}, is to consider the Mellin amplitude of one-loop supergravity correlators. In~\cite{Alday:2018kkw}, an ansatz in terms of an infinite double-sum of simultaneous double poles in the Mellin variables was proposed for $\me_2^{(2,0)}$. Recently, this was generalised to the $\corr22pp$ family of correlators, whose Mellin amplitudes take the form~\cite{Alday:2019nin}\footnote{The sums are divergent in the form written above and one should use a regularisation scheme, e.g. a zeta function regularisation. See also the recent reference~\cite{Aprile:2020luw} for a finite form of the one-loop supergravity Mellin amplitudes.}
\begin{align}\label{eq:one-loop_sugra_amplitude}
\me_p^{(2,0)} &= \sum_{n,m\geq0} \biggl[\frac{c^{u}_{mn}}{(s-4-2m)(t-(2+p)-2n)}+\frac{c^{t}_{mn}}{(s-4-2m)(\tilde{u}-(2+p)-2n)}\notag \\
&\qquad \qquad +\frac{c^{s}_{mn}}{(t-(2+p)-2n)(\tilde{u}-(2+p)-2m)}\biggr]_{\text{reg}}\, ,
\end{align}
with $c^{u}_{mn}=c^{t}_{mn}$ due to crossing symmetry. The residues $c^u_{mn}$ and $c^s_{mn}$ can then be fixed by matching against the position space leading logarithmic singularity in the two distinct crossing orientations of the $\corr22pp$ correlator, which can be directly obtained by exploiting the hidden ten-dimensional conformal symmetry from~\cite{Caron-Huot:2018kta}. In fact, we will use a similar reasoning in the next section to obtain the first string correction to the one-loop leading logarithmic singularity at order $\ll$.

Interestingly, it turns out that any additional single poles are absent in the above Mellin amplitude. In particular, there are no extra poles in what we call the \textit{window}-region, i.e. the region with $s=4,6,\ldots,2p-2$, where the gamma-functions $\Gamma(\frac{4-s}{2})\Gamma(\frac{2p-s}{2})$ from equation~\eqref{eq:Mellin_definition} do not overlap and thus produce only single poles. This region corresponds to twists $4\leq\tau<2p$ where the three-point functions $C_{pp,i}$ are subleading and thus there is no contribution to the leading logarithmic singularity, which is consistent with at most double-poles in the Mellin amplitude. It is non-trivial that the OPE predictions in the window from subleading three-point functions are already fully captured by the Mellin amplitude~\eqref{eq:one-loop_sugra_amplitude}, rendering any extra window poles obsolete. In fact, we will argue that this feature is not shared by one-loop string amplitudes, where an à priori unbounded number of additional window poles is necessary to fit the OPE predictions in the window.

\subsection{One-loop $\ll$}
Compared to the one-loop supergravity Mellin amplitudes, where the infinite double-sums require regularisation, a finite Mellin amplitude has been proposed for the tower of one-loop string-corrections in~\cite{Alday:2018kkw}.\footnote{See also~\cite{Drummond:2019hel} for the corresponding position space results.} The result for $\me_2^{(2,3)}$ suggests a natural generalisation to the $\corr22pp$ family of correlators:
\begin{align}\label{eq:one_loop_mellin_ansatz}
	\mathcal{M}^{(2,3)}_{p} = f_p(s) \, \widetilde{\psi}_0\Big{(}\frac{4-s}{2}\Big{)} + g_p(t)\, \widetilde{\psi}_0\Big{(}\frac{p+2-t}{2}\Big{)} + g_p(\tilde{u})\, \widetilde{\psi}_0\Big{(}\frac{p+2-\tilde{u}}{2}\Big{)} + R_p,
\end{align}
where $f_p$ and $g_p$ are fourth order polynomials which depend only on a single Mellin variable due to the truncation of the tree-level anomalous dimensions $\eta^{(1,3)}_i$ to spin zero. As a consequence of crossing symmetry, the $t$ and $\tilde{u}$-terms are given by the same polynomial $g_p$. Note that for convenience we use the digamma function shifted by the Euler-Mascheroni constant: $\widetilde{\psi}_0(z)\equiv\psi_0(z)+\gamma_E$.

We will first describe how to obtain the necessary one-loop double discontinuities, from which we can then determine the polynomials $f_p$ and $g_p$, thereby fixing the one-loop amplitudes $\me_p^{(2,3)}$ for twists $\tau\geq 2p$. However, as discussed earlier, we need to allow for possible additional poles in the window-region contributing to twists $\tau=4,6,\ldots,2p-2$, which are denoted by $R_p$ in~\eqref{eq:one_loop_mellin_ansatz}. We will discuss the computation of these extra window poles from subleading OPE predictions in detail in Section~\ref{sec:window_poles}, where we also describe the resolution of the associated mixing of subleading three-point functions.

\subsubsection{One-loop double discontinuities in position space}
At any order in the large $N$, large $\lambda$ double-expansion, the leading logarithmic singularity is completely determined by tree-level data. In the present case, we are interested in the one-loop double discontinuity at order $\ll$, which in position space is given by the expansion
\begin{align}\label{eq:one_loop_ddisc}
	\mathcal{H}_p^{(2,3)}|_{\text{log}^2(u)}=\sum_{t,l}{\sum_{i=1}^{t-1}{C^{(0,0)}_{22,i} C^{(0,0)}_{pp,i}} \eta^{(1,0)}_{i} \eta^{(1,3)}_{i}} G_{t,\ell}(x,\bar{x}).
\end{align}
The tree-level supergravity anomalous dimensions $\eta^{(1,0)}_{i}$ have been studied in~\cite{Aprile:2017xsp_unmixing,Aprile:2018efk}. For convenience, let us repeat the general formula for all $su(4)$ representations $[a,b,a]$:
\begin{align}
	\eta^{(1,0)}|_{[a,b,a]} =  -\frac{2M^{(4)}_t M^{(4)}_{t+\ell+1}}{\left(\ell+2(i+r)+a-\frac{1+(-1)^{a+\ell}}{2}\right)_6},
\label{eq:anom_dim_all_channels}
\end{align}
where $M_t^{(4)} = (t-1)(t+a)(t+a+b+1)(t+2a+b+2)$, the twist $\tau$ is parametrised by $t=\frac{\tau-b}{2}-a$, and $(i,r)$ are degeneracy labels given by
\begin{align}
	i=1,\ldots,(t-1), \qquad r=0,\ldots,(\mu-1), \qquad \mu=\left\{\begin{array}{ll}
	\bigl\lfloor{\frac{b+2}2}\bigr\rfloor \quad &a+l \text{ even,}\\[.2cm]
	\bigl\lfloor{\frac{b+1}2}\bigr\rfloor \quad &a+l \text{ odd.}
	\end{array}\right.
\end{align}
Further string corrections to the double-trace spectrum have been addressed in~\cite{Drummond:2019odu}. An important aspect of the spectrum at order $\ll$ is that only operators in the $[0,b,0]$ representation with degeneracy labels $(i,r)=(1,0)$ and spin $\ell=0$ acquire a correction.\footnote{As argued in~\cite{Drummond:2019odu}, this is a consequence of the effective ten-dimensional spin $\ell_{10}$ being constrained to the value $\ell_{10}=0$ for the present case of a $\mathcal{R}^4$ contact interaction vertex in the string theory effective action.} Their anomalous dimensions read~\cite{Drummond:2019odu,Drummond:2020dwr}
\begin{align}\label{eq:anom_dim_-3/2}
	\eta^{(1,3)}_{i=1} = -\frac{\zeta_3}{840}M^{(4)}_t M^{(4)}_{t+\ell+1}(t-1)_3(t+b+1)_3\cdot\delta_{\ell,0}.	
\end{align}
The truncation to spin zero and $i=1$ simplifies the sum~\eqref{eq:one_loop_ddisc} drastically, which now becomes
\begin{align}\label{eq:one_loop_ddisc_simplify}
	\mathcal{H}_p^{(2,3)}|_{\text{log}^2(u)}=\sum_{t}{C^{(0,0)}_{22,1} C^{(0,0)}_{pp,1}} \eta_1^{(1,0)} \eta_1^{(1,3)}  G_{t,\ell=0}(x,\bar{x}),
\end{align}
where all factors are understood to have $\ell=0$. The fact that only the $i=1$ state contributes to the sum allows for a particularly efficient route to compute the double discontinuities, avoiding a direct resummation of the above sum altogether. Let us describe this convenient shortcut next.

It was noticed in~\cite{Alday:2017vkk,Aprile:2017qoy_2233,Aprile:2018efk,Caron-Huot:2018kta,Aprile:2019rep} that the use of differential operators constructed from certain quadratic and quartic Casimirs (of which the conformal blocks are eigenfunctions) may simplify expressions such as the above~\eqref{eq:one_loop_ddisc_simplify}. In particular, in~\cite{Aprile:2017qoy_2233} an eighth-order differential operator $\Delta^{(8)}$ was constructed, whose action on the conformal block exactly produces the numerator of the supergravity anomalous dimensions~\eqref{eq:anom_dim_all_channels}, and it was shown that it is beneficial to pull out this operator from the supergravity double discontinuities. Generalised to account for the internal $su(4)$ variables, $\Delta^{(8)}$ takes the form~\cite{Caron-Huot:2018kta}
\begin{align}
\Delta^{(8)} = \frac{x\bar{x}y\bar{y}}{(x-\bar{x})(y-\bar{y})} \prod_{i,j=1}^2\Big( \mathbf{C}^{[\alpha,\beta,0]}_{x_i}-\mathbf{C}^{[-\alpha,-\beta,0]}_{y_j}\Big) \frac{(x-\bar{x})(y-\bar{y})}{x\bar{x}y\bar{y}},
\end{align}
where for a correlator with general external charges $p_i$ we have introduced $\alpha=(p_2-p_1)/2$, $\beta=(p_3-p_4)/2$, and
\begin{align}
\mathbf{C}^{[\alpha,\beta,\gamma]}_{x} = x^2 (1-x) \partial^{2}_{x} + x(\gamma-(1+\alpha+\beta)x)\partial_x-\alpha\beta x.
\end{align}
For the correlators $\langle \mathcal{O}_2 \mathcal{O}_2 \mathcal{O}_p \mathcal{O}_p \rangle$, where only the singlet $su(4)$ representation is exchanged, there is no $y$ or $\bar{y}$ dependence in the function on which $\Delta^{(8)}$ acts. The operator then reduces to\footnote{This operator, or one related to it by a simple crossing transformation, appeared in \cite{Drummond:2006by} as a way of relating the superconformal primary components to the axion-dilaton components of the stress-energy four-point function.} 
\be
\Delta^{(8)}_{[0,0,0]} = \frac{x^4 \bar{x}^4}{x-\bar{x}} \bigl[\partial_x^2 \partial_{\bar{x}}^2 (1-x)^2(1-\bar{x})^2 \partial_x^2 \partial_{\bar{x}}^2 \bigr] (x-\bar{x})\,.
\label{Delta8singlet}
\ee
Now, thanks to the truncation of the sum~\eqref{eq:one_loop_ddisc_simplify} to $i=1$ only, pulling out $\Delta^{(8)}$ amounts to simply removing the supergravity anomalous dimension from the sum (up to an overall factor from its denominator), we arrive at
\begin{align}\label{eq:ddisc_delta8}
	\mathcal{H}_p^{(2,3)}|_{\log^2(u)}&=-\frac{1}{360} u^{-2} \Delta^{(8)}_{[0,0,0]} \Bigl[ u^2 \sum_{t}{C^{(0,0)}_{22,1} C^{(0,0)}_{pp,1}} \eta_1^{(1,3)}  G_{t,\ell=0}(x,\bar{x})\Bigr]\,.
\end{align}
Finally, we recognise that the remaining sum is nothing else than the single discontinuity of the tree-level correlator $\mathcal{H}^{(1,3)}_p$,
\be
\mathcal{H}_p^{(1,3)}|_{\log(u)} = \sum_{t}{C^{(0,0)}_{22,1} C^{(0,0)}_{pp,1}} \eta_1^{(1,3)}  G_{t,\ell=0}(x,\bar{x})\,.
\ee
This result for the double discontinuity can be straightforwardly generalised to all correlators with arbitrary external charges, resulting in the following simple relation,
\begin{align}\label{eq:ddisc_position_space}
\mathcal{H}^{(2,3)}_{p_1p_2p_3p_4}|_{\log^2(u)}=-\frac{1}{360}u^{-2}(u \sigma)^{-\frac{p_{43}}{2}}\Delta^{(8)}\big[u^2(u \sigma)^{\frac{p_{43}}{2}}\mathcal{H}_{p_1p_2p_3p_4}^{(1,3)}|_{\log(u)}\big].
\end{align}
In the case of correlators $\langle \mathcal{O}_2 \mathcal{O}_p \mathcal{O}_2 \mathcal{O}_p \rangle$ the above expression reduces to
\be
\mathcal{H}^{(2,3)}_{2p2p}|_{\log^2(u)}=-\frac{1}{360}u^{-2}\Delta^{(8)}_{[0,p-2,0]} \big[u^2 \mathcal{H}_{2p2p}^{(1,3)}|_{\log(u)}\big]
\ee
where
\be
\Delta^{(8)}_{[0,b,0]} = \frac{x^4 \bar{x}^4}{x-\bar{x}} \bigl[\partial_x^2 \partial_{\bar{x}}^2 (x \bar{x})^{-b} (1-x)^2(1-\bar{x})^2 \partial_x^2 \partial_{\bar{x}}^2 \bigr](x \bar{x})^{b} (x-\bar{x})\,
\ee
generalises the operator (\ref{Delta8singlet}).

The computation of one-loop double discontinuities at order $\ll$ is thus reduced to acting with $\Delta^{(8)}$ on the $\log(u)$ part of tree-level correlators. Note that these correlators are fully determined through the flat space limit, and the result for arbitrary external charges is given by~\cite{Drummond:2019odu}
\begin{align}\label{eq:tree-level_result_-3/2}
	\mathcal{H}^{(1,3)}_{p_1p_2p_3p_4} = \frac{(\Sigma-1)_3 \zeta_3}{4} B^{\text{sugra}}_{p_i}(\sigma,\tau) u^{\frac{p_1+p_2+p_3-p_4}{2}} \dbar{p_1+2,p_2+2,p_3+2,p_4+2},
\end{align}
with $B^{\text{sugra}}_{p_i}(\sigma,\tau)$ defined in~\eqref{eq:B_sugra} and $\Sigma$ half the sum of external charges, $\Sigma=\frac{p_1+p_2+p_3+p_4}{2}$.

Lastly, let us note that the above result for $\mathcal{H}^{(1,3)}_{p_1p_2p_3p_4}$ can also be obtained from a ten-dimensional generating functional, along the lines of the hidden ten-dimensional conformal symmetry discovered for tree-level supergravity correlators~\cite{Caron-Huot:2018kta}. In the supergravity case, a ten-dimensional generating functional can be used to define the differential operator $\mathcal{D}_{p_1p_2p_3p_4}$, which generates correlators of arbitrary external charges by application to the seed-correlator $\fourtwo$, i.e.
\begin{align}
	\mathcal{H}^{(1,0)}_{p_1p_2p_3p_4} \propto \mathcal{D}_{p_1p_2p_3p_4}~u^4\dbar{2422}(u,v).
\end{align}
Due to the simplicity of the double-trace spectrum at order $a\ll$ (or in other words, because of the simplicity of the Mellin amplitude $\me^{(1,3)}$) the same construction from~\cite{Caron-Huot:2018kta} applies to the first string correction and we find
\begin{align}
	\mathcal{H}^{(1,3)}_{p_1p_2p_3p_4} \propto \mathcal{D}_{p_1p_2p_3p_4}~u^4\dbar{4444}(u,v),
\end{align}
which repackages all correlators to descend from the seed $u^2\mathcal{H}^{(1,3)}_2\propto u^4\dbar{4444}$.

To summarise, we have arrived at the main formula~\eqref{eq:ddisc_position_space} for the one-loop double discontinuity, given by $\Delta^{(8)}$ on the known tree-level $\ll$ correlator~\eqref{eq:tree-level_result_-3/2}, which in turn can be obtained from a ten-dimensional generating function analogous to the supergravity case discussed in~\cite{Caron-Huot:2018kta}. In the remainder of this section we will discuss the conversion of the double discontinuities into Mellin space.
\subsubsection{Matching the double discontinuity from Mellin space}
The one-loop double discontinuities for the $\corr22pp$ correlators as computed from~\eqref{eq:ddisc_position_space} are of the form
\begin{align}
	\mathcal{H}_{p}^{(2,3)}|_{\log^2(u)} \sim \frac{a_1(x,\bar{x})+a_2(x,\bar{x})[\log(1-x)-\log(1-\bar{x})]}{(x-\bar{x})^{17+2p}},
\end{align}
where $a_1$, $a_2$ are polynomials with their degree bounded by the denominator power $17+2p$. In order to facilitate the comparison with the small $(u,v)$ expansion of the Mellin space amplitudes, we change to the variables $(x,\tilde{x})=(x,1-\xb)$ which are related to the usual conformal cross-ratios by $u=x \bar{x} =x(1-\tilde{x})$ and $v=(1-x)(1-\bar{x})=(1-x)\tilde{x}$.

We have now all the necessary ingredients to determine the polynomials $f_p$ and $g_p$ in our Mellin space ansatz for $\me^{(2,3)}_p$ in equation~\eqref{eq:one_loop_mellin_ansatz}: firstly, we consider the $\log^2(u)\log(v)$ contribution which arises from triple-poles in $s$ and double-poles in $t$. As only the first term in~\eqref{eq:one_loop_mellin_ansatz} contributes, matching it against the corresponding term in the double discontinuity computed from~\eqref{eq:ddisc_position_space} fully determines the polynomial $f_p(s)$, which is found to be consistent with the formula,
\begin{align}\label{eq:f_p}
\begin{split}
	f_p(s) &= -\frac{p(p)_4\zeta_3}{360(p-2)!}\Big((p+4)_4s^4-4(p+4)_3(7p+9)s^3\\
		   &\qquad\qquad\qquad +4(p+4)_2(71p^2+203p+102)s^2\\
		   &\qquad\qquad\quad~~ -16(p+4)(77p^3+346p^2+399p+90)s+1920(p)_4\Big).
\end{split}
\end{align}
In a second step, in order to determine the polynomial $g_p$, we cross the correlator from the $\corr22pp$ orientation to $\langle\ot\op\ot\op\rangle$, which in Mellin space corresponds to the exchange of $s$ and $\tilde{u}$. In this orientation, matching the double discontinuity fixes the polynomial $g_p$. We find
\begin{align}\label{eq:g_p}
\begin{split}
	g_p(t) &= f_p(t)+\frac{(p-2)p(p)_4\zeta_3}{360(p-2)!}\Big(4p(p+4)_3t^3-6(p+4)_2(p^3+11p^2+14p-12)t^2\\
		   &\qquad\qquad\qquad\qquad\qquad~~ +4(p+4)(p^5+14 p^4+106 p^3+239 p^2-6 p-252)t\\
		   &\qquad\qquad\qquad\qquad\qquad~~~ -(p+2)_2(p^5+11p^4+86p^3+472p^2-96p-576)\Big).
\end{split}
\end{align}
Note that for $p=2$ we have $f_2(s)=g_2(s)$ (consistent with the enhanced crossing symmetry of the $\fourtwo$ correlator) and the result agrees with $\me_2^{(2,3)}$ previously found in~\cite{Alday:2018kkw,Drummond:2019hel}.

At this stage, we have determined the Mellin amplitude $\me_p^{(2,3)}$ above the window-region, i.e. for twists $\tau\geq2p$. We will now turn our attention to the window-region and discuss how the additional single poles $R_p$ can be fixed by unmixing the subleading three-point functions in the window.
\section{Determining the window poles}\setcounter{equation}{0}\label{sec:window_poles}
Having fixed the polynomials $f_p(s)$ and $g_p(t)$ in the ansatz (\ref{eq:one_loop_mellin_ansatz}), we now consider the window contributions. The window has been defined as the region in which the poles of the gamma-funcions $\Gamma(\frac{4-s}{2})\Gamma(\frac{2p-s}{2})$ from (\ref{eq:Mellin_definition}) are non-overlapping. Therefore, in Mellin space, these terms are given by at most double poles in s. This suggests including a series of poles, denoted $R_p$ in our ansatz (\ref{eq:one_loop_mellin_ansatz}), taking the form 
\begin{align}
 	R_{p}=\sum_{i=2}^{p-1} \frac{\alpha_{i}(p)}{s-2i}.
\label{eq:ansatz_window_poles}
\end{align}
The residues $\alpha_i(p)$ are then fixed by matching window contributions to the logarithmic part of the correlator, which is predicted by OPE data. The full space-time expansion of the one-loop $\log(u)$ contribution is given by
\begin{align}
	\mathcal{H}_p^{(2,3)}|_{\log(u)} = \sum_{t,l}\sum_{i}^{t-1} & \bigg{[}\left(C^{(0,0)}_{22,i}C^{(1,3)}_{pp,i} +C^{(1,3)}_{22,i}C^{(0,0)}_{pp,i}\right)\eta^{(1,0)}_{i} + \left(C^{(0,0)}_{22,i}C^{(1,0)}_{pp,i} + C^{(1,0)}_{22,i}C^{(0,0)}_{pp,i}\right)\eta^{(1,3)}_{i} \notag \\
	& + C^{(0,0)}_{22,i}C^{(0,0)}_{pp,i}\eta^{(2,3)}_{i}\bigg{]} G_{t,l} + 2 C_{22,i}^{(0,0)}C_{pp,i}^{(0,0)}\eta^{(1,0)}_{i}\eta^{(1,3)}_{i}\, \nabla_t  G_{t,l},
\label{eq:one_loop_ope_full}
\end{align}
where $\nabla_t \equiv u^t \partial_t u^{-t}$. Let us recall the definition of the window-region for correlators of the more general form $\langle ppqq\rangle$. It is useful to first recap some features of three-point functions set out in \cite{Aprile:2019rep}. Firstly, the leading order three-point functions $C_{pp,i}^{(0,0)}$ in~\eqref{eq:c_expansions} are only non-vanishing for twists $\tau \geq 2p$. Hence, when considering products such as $C_{pp,i}\,C_{qq,i}$, there is a region of twists within which only one of the leading three-point functions is switched on, with the other one vanishing. It is this region which we call the window for the $\langle ppqq \rangle$ correlators, and is given by $p \leq t < q$. In the small $u$ expansion the powers $u^2,\ldots,u^{p-1}$ are determined only by the window-region.

When restricting to the window-region for $\langle 22pp \rangle$ (where $t=2,3,...,p-1$) the leading order three-point functions $C^{(0,0)}_{pp,i}$ vanish. With four of the terms now absent, \eqref{eq:one_loop_ope_full} becomes
\begin{align}
	\sum_{\ell}\sum_{i=1}^{t-1}{C^{(0,0)}_{22,i}\Big{(} C^{(1,0)}_{pp,i} \eta^{(1,3)}_i  +  C^{(1,3)}_{pp,i} \eta^{(1,0)}_i}\Big{)} G_{t,l}(u,v),
	\label{eq:window_log_ope}
\end{align}
where $C^{(1,k)}_{pp,1}$ are the tree-level supergravity and string corrected three-point functions for $k=0$ and $k=3$ respectively. The three-point functions appearing in (\ref{eq:window_log_ope}) can be extracted from the non-$\log(u)$ contribution to tree-level correlators of the form $\langle ppqq \rangle$. The generalisation to the set of correlators $\langle ppqq \rangle$ is essential for the unmixing of degenerate operators, which we detail in the following.

\subsection{Unmixing}
As indicated in (\ref{eq:window_log_ope}) there is not a one-to-one correspondence between three-point functions and conformal block coefficients. Therefore, to calculate the individual three-point functions we need to unmix the degenerate operators entering the block coefficients. The correlators $\langle 22pp \rangle$ do not provide enough information to solve this degeneracy problem, instead we must consider a more general set of correlators taking the form $\langle ppqq \rangle$. At each level in twist we have $(t-1)$ degenerate operators, thus to solve we must consider the set of $(t-1)$ families of correlators with $2 \leq p \leq t$. As mentioned before, the relevant information is encoded in the non-$\log(u)$ contribution to the tree-level correlators, which have the space-time expansion 
\begin{align}
	\mathcal{H}_{ppqq}^{(1,k)}|_{\text{non-log(u)}}=\sum_{t,\ell}\sum_{i=1}^{t-1} \left(C^{(0,0)}_{pp,i} C^{(1,k)}_{qq,i}+C^{(1,k)}_{pp,i} C^{(0,0)}_{qq,i}\right) G_{t,\ell} + C^{(0,0)}_{pp,i}C^{(0,0)}_{qq,i}\eta_i^{(1,k)}\,\nabla_t G_{t,l}\,.
	\label{eq:non_log_ope_ppqq}
\end{align}
When looking at the window-region ($p \leq t < q$), the only surviving term from~\eqref{eq:non_log_ope_ppqq} is given by
\begin{align}
	\sum_{\ell}{\sum_{i=1}^{t-1}{C^{(0,0)}_{pp,i} C^{(1,k)}_{qq,i}}}\, G_{t,\ell}(u,v).
	\label{eq:non_log_ope_ppqq_window}
\end{align}
Having detailed where the required data can be found, the unmixing procedure is best illustrated with an example. With one operator at twist four, and therefore no mixing, the three-point functions can indeed be calculated just using data from the $\langle 22qq \rangle$ family. Thus, the first instructive case where operator mixing happens is at twist six.\\
\subsubsection{Example: unmixing at twist six}
At twist six we wish to compute the couplings $C^{(1,k)}_{qq,1}$ and $C^{(1,k)}_{qq,2}$ for $k=0$ and $k=3$ (supergravity and string corrected) respectively. Following the discussion above, to have enough information to perform the unmixing both the $\langle 22qq \rangle$ and $\langle 33qq \rangle$ family of correlators are needed. To ensure twist six lies within the window for both sets of correlators, we must have $q>3$. As shown in~\eqref{eq:non_log_ope_ppqq_window}, within the window-region the conformal block coefficients $L_{2,\tau=6}^{(1,k)}$ and $L_{3,\tau=6}^{(1,k)}$ are given by
\begin{align}
\begin{split}
	C^{(0,0)}_{22,1} C^{(1,k)}_{qq,1}+C^{(0,0)}_{22,2} C^{(1,k)}_{qq,2} = L_{2,\tau=6}^{(1,k)}, \\
	C^{(0,0)}_{33,1} C^{(1,k)}_{qq,1}+C^{(0,0)}_{33,2} C^{(1,k)}_{qq,2} = L_{3,\tau=6}^{(1,k)},
\end{split}
\end{align}
for $\langle 22qq \rangle$ and $\langle 33qq \rangle$ respectively. This can be nicely repackaged in matrix form by 
\begin{gather}
 \begin{bmatrix} C^{(0,0)}_{22,1} & C^{(0,0)}_{22,2} \\ C^{(0,0)}_{33,1} & C^{(0,0)}_{33,2} \end{bmatrix}{}
 \begin{bmatrix} C^{(1,k)}_{qq,1} \\ C^{(1,k)}_{qq,2} \end{bmatrix}
 =
 \begin{bmatrix} L_{2,\tau=6}^{(1,k)} \\ L_{3,\tau=6}^{(1,k)} \end{bmatrix},
\end{gather}
from which the desired couplings can be readily obtained. This can be easily generalised to arbitrary twists
\begin{gather}
 \mathbb{C}_t^{(0,0)} \vec{\mathbf{C}}^{(1,k)}_t = \vec{\mathbf{L}}^{(1,k)}_t, \\
 (\mathbb{C}_t^{(0,0)})^{-1} \vec{\mathbf{L}}^{(1,k)}_t = \vec{\mathbf{C}}^{(1,k)}_t,
\end{gather}
where the matrix is now $(t-1) \times (t-1)$ dimensional. During this process much new OPE data has been generated, see Appendix~\ref{ap:OPE} for more details.
\subsection{Results for the window poles}
With all necessary OPE data at hand we can proceed, twist by twist, in calculating the residues $\alpha_i(p)$ in our ansatz. The window contribution is found to be
\begin{align}\label{eq:window_poles}
\begin{split}
	R_p&=\frac{16p(p)_3}{(p-2)!}\Big(\frac{8(p-2)}{s-4}+\frac{16(p-3)_2}{s-6}+\frac{8(p-4)_3}{s-8}+\frac{4}{3}\frac{(p-5)_4}{s-10}+\frac{1}{15}\frac{(p-6)_5}{s-12}\Big)\\[0.1cm]
	&=\frac{16p(p)_3}{(p-2)!} \sum_{n} \frac{192}{(n-1)!(n-2)!(6-n)!}~\frac{(p-n)_{n-1}}{s-2n}.
\end{split}
\end{align}
A few features of this result are worth mentioning. Firstly, the residues are non-zero. This means that simply extending the above-window poles (i.e. contributions to twists $\tau\geq2p$, captured by $f_p(s)$) down into the window-region does not entirely account for the operator-mixing at the level of subleading three-point functions, and thus it does not directly yield the correct one-loop amplitude. This should be put in contrast with the case of one-loop supergravity, see Section~\ref{sec:one_loop_sugra}, which does not seem to require any additional single poles in order to match the OPE predictions in the window~\cite{Alday:2019nin}.

Secondly, there are only five non-vanishing terms in the above sum. Recall that our ansatz from equation~(\ref{eq:ansatz_window_poles}) allowed for an arbitrarily large set of poles, growing linearly with increasing $p$. We find it highly non-trivial that the sum truncates and, in particular, that this finite number of poles correctly takes into account the entire OPE data in the window for all twists. Currently, we do not have any argument why precisely five poles are enough to accomplish this.

Lastly, note that each term corresponds to a tree-level s-channel exchange-diagram of an operator with twist at the double-trace location $\tau=4,6,\ldots,12$, respectively.\footnote{In position space, this simply evaluates to a linear combination of so-called $\bar{D}$-functions.} The presence of such tree-level correction-terms emphasises the fact that the knowledge of only the $\log^2(u)$-term is not sufficient to reconstruct the full one-loop correlator. Instead, an additional understanding of the physics in the window-region (and for correlators of more general external charges, similarly in the below-window region) is crucial, as already stressed in~\cite{Aprile:2019rep}. Let us now turn to the possibility of adding any tree-level contact-diagrams, which will show up as polynomial ambiguities in our Mellin space amplitudes.

\subsection{Polynomial ambiguities}\label{sec:ambiguities}
The OPE predictions for the double discontinuity and the window-region allow for the addition of in principle any polynomial of the Mellin variables. We will collectively refer to these polynomial terms as ambiguities, as they are not determined by any OPE consistency requirements. Note that these terms are of tree-level like form and correspond to the genus-one contributions to the modular completions of the tree-level string corrections $\l^{-\frac{k}{2}}\mathcal{H}^{(1,k)}$. In particular, the one-loop ambiguities at order $\ll$ are the modular completion of the tree-level $\partial^8\mathcal{R}^4$ term at order $\l^{-\frac{7}{2}}$, and as such we expect them to be polynomials of maximal degree four. They will therefore contribute only to finite spin in the superconformal block decomposition, i.e. up to spin four.

In order to fix these ambiguities, we have to rely on different methods. One possibility is to consider the flat space limit (see Section~\ref{sec:flat_space_limit} for more details), which due to the vanishing quartic contribution in~\eqref{eq:genus-one_analytic} in the analytic part of the genus-one string amplitude sets the degree-four ambiguities to zero. Presently, we do not have another method on how to fix the remaining ambiguities of up to cubic degree. A possible method might be given by supersymmetric localisation techniques, which were able to fix a similar ambiguity in the one-loop supergravity correlator, see~\cite{Chester:2019pvm,Chester:2020dja}.

\section{Towards higher charges: the $\langle\ott\ott\ott\ott\rangle$ correlator}\setcounter{equation}{0}\label{sec:3333}
Let us here present a first extension of the above results to correlators of the next degree in extremality, where we will encounter non-trivial dependence on the R-symmetry cross-ratios. The simplest such correlator is $\langle\ott\ott\ott\ott\rangle$. Being a polynomial in $\sigma$ and $\tau$ of degree one, there are three $su(4)$ R-symmetry channels: $[0,0,0]$, $[1,0,1]$ and $[0,2,0]$.\footnote{The three $su(4)$ channels are connected to the R-symmetry cross-ratios via the harmonic polynomials
\begin{align}\label{eq:su(4)_channels_3333}
	Y_{[0,0,0]}=1,\qquad Y_{[1,0,1]}=\sigma-\tau,\qquad	Y_{[0,2,0]}=\tfrac{1}{2}(\sigma+\tau)-\tfrac{1}{6}.
\end{align}}

We start with a manifestly fully crossing symmetric ansatz in Mellin space:\footnote{Note that for the case of the $\langle\ott\ott\ott\ott\rangle$ correlator the string of six gamma-functions in the definition of the Mellin transform~\eqref{eq:Mellin_definition} reads
\begin{align}\label{eq:gammas_3333}
	\Gamma_{3333}\equiv\Gamma^2\Big(\frac{6-s}{2}\Big)\Gamma^2\Big(\frac{6-t}{2}\Big)\Gamma^2\Big(\frac{6-\tilde{u}}{2}\Big).
\end{align}}
\begin{equation}\label{eq:3333_mellin_amplitude}
\me_{3333}^{(2,3)}= h(s;\sigma,\tau) \widetilde{\psi}_0\Big(\frac{6-s}{2}\Big) + \tau h(t;\tfrac{\sigma}{\tau},\tfrac{1}{\tau}) \widetilde{\psi}_0\Big(\frac{6-t}{2}\Big) + \sigma h(\tilde{u};\tfrac{1}{\sigma},\tfrac{\tau}{\sigma}) \widetilde{\psi}_0\Big(\frac{6-\tilde{u}}{2}\Big) + R_{3333},
\end{equation}
where $h(s;\sigma,\tau)$ is a polynomial of degree one in the R-symmetry cross-ratios and, as before, a fourth order polynomial in the Mellin variables. Potential additional single poles are denoted by $R_{3333}$. 

We proceed as in the previous cases and compute first the one-loop double discontinuity using formula~\eqref{eq:ddisc_position_space}, which we then match against the above Mellin space ansatz. We find
\begin{align}
\begin{split}
	h(s;\sigma,\tau) & = -1701\zeta_3\big((55 s^4-740 s^3+4172 s^2-11360 s+12368)\\
					 &\qquad\qquad+(55 s^4-900 s^3+5804 s^2-17440 s+20496)(\sigma+\tau)\big).
\end{split}
\end{align}
Note that $h(s;\sigma,\tau)$ contributes only to the singlet and $[0,2,0]$ channel, and is thus symmetric in $\sigma$ and $\tau$. This is a consequence of the simplicity of the string anomalous dimensions $\eta^{(1,3)}$ which vanish in channels $[a,b,a]$ with $a\neq0$. Furthermore, let us remark that the $s^4$ coefficient of $h(s;\sigma,\tau)$ is proportional to the factor $(1+\sigma+\tau)$, which is necessary for matching the flat space limit as we will describe in Section~\ref{sec:flat_space_limit}.

As a second step, we need to consider the additional single poles $R_{3333}$. Since we have already fixed the correlator for twists $\tau\geq6$ in all three channels, we are left with a potential twist four contribution, which, due to the higher unitarity bound in the $[1,0,1]$ and $[0,2,0]$ channels, can appear only in the singlet.\footnote{For a given $su(4)$ channel $[a,b,a]$, the unitarity bound is given by $2a+b+2$ and therefore the genuine long supermultiplets necessarily have twists $\tau>2a+b+2$.} A fully crossing symmetric ansatz for $R_{3333}$, contributing to twist four only in the singlet channel, reads
\begin{align}
	R_{3333}(s,t;\sigma,\tau)=\beta\,\Big(\frac{1}{s-4}+\frac{\tau}{t-4}+\frac{\sigma}{\tilde{u}-4}\Big).
\end{align}
To determine $\beta$, let us consider the OPE  prediction for the singlet channel at twist 4. In contrast to the previous discussion, this twist 4 singlet contribution lies \textit{below the window}, as the $s=4$ pole in $R_{3333}$ does not overlap with any of the $s$-poles in the gamma-functions~\eqref{eq:gammas_3333}. This thus corresponds to an analytic contribution to the correlator, for which the OPE at twist 4 ($t=2$) gives the expression
\begin{align}
	\sum_{\ell} 2\,C^{(1,0)}_{33,1} C^{(1,3)}_{33,1}\, G_{t=2,\ell}(u,v),
\end{align}
where $C_{33,1}^{(1,0)}$ and $C_{33,1}^{(1,3)}$ can be found in Appendix~\ref{ap:OPE}. Matching the above OPE prediction determines $\beta$ to take the value
\begin{align}\label{eq:beta_value}
	\beta=-1036800\zeta_3,
\end{align}
leaving us with a fully fixed Mellin amplitude $\me_{3333}^{(2,3)}$ (up to the usual set of polynomial ambiguities described in Section~\ref{sec:ambiguities}).

This result initiates the study of correlators with general external charges, and we believe that the previously found property of a truncated number of extra window poles will generalise. Note that generic higher charge correlators have both a non-empty window and below-window region, each of which has a different OPE origin and therefore has to be supplemented with its own tower of single poles. We leave this more general problem for future investigations.

\section{New twist 5 and 6 one-loop anomalous dimensions}\setcounter{equation}{0}\label{sec:twist5_anom_dim}
With the Mellin amplitudes $\me_p^{(2,3)}$ and $\me_{3333}^{(2,3)}$ at hand, we can now use them to extract new subleading CFT data at this order. However, as discussed previously, for general twists one has to solve a mixing problem as there are many degenerate double-trace operators. Only for specific $su(4)$ channels at the lowest twist there is a unique double-trace operator whose anomalous dimension one can straightforwardly extract, with the first few cases being the singlet channel at twist four, the $[0,1,0]$ channel at twist five and the $[1,0,1]$ channel at twist six.

The singlet channel twist four anomalous dimension at order $a^2\ll$ can be extracted from $\me_2^{(2,3)}$ and was already given in~\cite{Binder:2019jwn}:
\begin{align}\label{eq:twist4_anom_dim}
	\eta^{(2,3)}_4|_{[0,0,0]} = -1658880\zeta_3\frac{(\ell+2)_4(\ell^2+7\ell+16)(\ell^2+7\ell+54)}{(\ell-4)_6(\ell+6)_6},~\text{ for even spins }\ell\geq6,
\end{align}
where the restriction on spin is due to the finite spin ambiguities which contribute up to spin four at this order.

Similarly, we can obtain the twist five anomalous dimension from our new result for $\me_3^{(2,3)}$. After crossing the $\langle\ot\ot\ott\ott\rangle$ correlator to the orientation $\langle\ot\ott\ot\ott\rangle$, we can access the twist five contribution in the $[0,1,0]$ channel. We find
\begin{align}\label{eq:twist5_anom_dims}
\begin{split}
	\eta^{(2,3)}_{5,\text{even}}|_{[0,1,0]} &= -5529600\zeta_3\frac{(\ell+2)_2(\ell+5)_2(\ell^3+14\ell^2+103\ell+280)}{(\ell-4)_6(\ell+7)_5}+\frac{20\alpha}{(\ell+1)(\ell+4)^2(\ell+7)},\\
	\eta^{(2,3)}_{5,\text{odd}}|_{[0,1,0]}  &= -5529600\zeta_3\frac{(\ell+2)_2(\ell+5)_2(\ell^3+10\ell^2+71\ell+160)}{(\ell-3)_5(\ell+7)_6}+\frac{20\alpha}{(\ell+1)(\ell+4)^2(\ell+7)},
\end{split}
\end{align}
which are valid for even spins $\ell\geq6$ and odd spins $\ell\geq5$, respectively. The second term arises from the extra window pole at $s=4$ and the value of $\alpha$ follows from the general formula~\eqref{eq:window_poles}, giving $\alpha=23040\zeta_3$.

Finally, from the projection of $\me_{3333}^{(2,3)}$ in~\eqref{eq:3333_mellin_amplitude} to the $[1,0,1]$ channel, we can extract information on twist six anomalous dimensions.\footnote{Note that even though the projection to the $[1,0,1]$ channel of $\me_{3333}^{(2,3)}$ by construction does not contribute to the $\log^2(u)$ term, the full crossing symmetry of the correlator implies a non-vanishing $\log(u)$ contribution from which we extract the one-loop anomalous dimension.} In principle, as well as the single double-trace operator, there can be triple-trace operators at twist six. If we assume that no triple-trace operators contribute, the twist six anomalous dimension of the double-trace operator reads
\begin{align}\label{eq:twist6_anom_dim}
\begin{split}
	\eta^{(2,3)}_6|_{[1,0,1]} &= -\frac{278691840\zeta_3 \left(13 \ell^6+351 \ell^5+3355 \ell^4+13005 \ell^3+20752 \ell^2+43884 \ell+22320\right)}{(\ell-3) (\ell-2) (\ell-1) \ell (\ell+1) (\ell+3) (\ell+6) (\ell+8) (\ell+9) (\ell+10) (\ell+11) (\ell+12)}\\
	&\quad+\frac{16\beta}{(\ell+1) (\ell+3) (\ell+6) (\ell+8)},~\text{ for odd spins }\ell\geq5,
\end{split}
\end{align}
where $\beta=-1036800\zeta_3$ as determined in~\eqref{eq:beta_value}.

In reference~\cite{Aprile:2017qoy_2233}, a non-trivial $\mathbb{Z}_2$ symmetry of the double-trace anomalous dimensions was observed. In particular, it was found that the supergravity anomalous dimensions $\eta^{(1,0)}$ and $\eta^{(2,0)}$ are symmetric under the reciprocity symmetry $\ell\rightarrow-\ell-n(t,a,b)$, where $n$ is an integer shift depending on the twist and $su(4)$ channel $[a,b,a]$. We find that the one-loop string corrected anomalous dimensions given above in~\eqref{eq:twist4_anom_dim},~\eqref{eq:twist5_anom_dims} and~\eqref{eq:twist6_anom_dim} continue to obey this symmetry. Indeed, at twists four and six one can check that
\begin{align}
\begin{split}
	[0,0,0]:\qquad\eta^{(2,3)}_4(-\ell-7) &= \eta^{(2,3)}_4(\ell),\\
	[1,0,1]:\qquad\eta^{(2,3)}_6(-\ell-9) &= \eta^{(2,3)}_6(\ell),
\end{split}
\end{align}
whereas at twist five the even and odd spin contributions map into each other
\begin{align}
	[0,1,0]:\qquad\eta^{(2,3)}_{5,\text{even}}(-\ell-8) = \eta^{(2,3)}_{5,\text{odd}}(\ell),\quad\eta^{(2,3)}_{5,\text{odd}}(-\ell-8) = \eta^{(2,3)}_{5,\text{even}}(\ell),
\end{align}
consistent with the discussion presented in~\cite{Aprile:2017qoy_2233}.

\section{The flat space limit}\setcounter{equation}{0}\label{sec:flat_space_limit}
In this section we first review the flat space limit of Mellin amplitudes, paying attention to keep the discussion general such that the formalism can be applied to correlators with arbitrary external charges. Only in the end we will specialise to the $\corr22pp$ family of correlators and verify that the Mellin amplitudes found in the previous sections agree with the low-energy expansion of the ten-dimensional type IIB amplitude, providing a non-trivial consistency check for our Mellin amplitudes.

\subsection{Review of the flat space limit for arbitrary charge correlators}
Let us start by reviewing the general flat space limit formula for four-particle Mellin amplitudes. A relation between Mellin amplitudes and scattering amplitudes in AdS was first motivated in~\cite{Penedones:2010ue}, and explored further in~\cite{Fitzpatrick:2011hu}. In four dimensions, the relation reads\footnote{Note that the above relation~\eqref{eq:Penedones_formula} requires the use of the full Mellin amplitude $M(s,t;\sigma,\tau)$ which is related to the \textit{reduced} Mellin amplitude $\me(s,t;\sigma,\tau)$ as defined in~\eqref{eq:Mellin_definition} through the action of a difference operator corresponding to the factor $\mathcal{I}$ in~\eqref{eq:superconformal_constraint}. In the flat space limit $s,t\rightarrow\infty$, this is given by
\begin{align}\label{eq:full_Mellin_amplitude}
	M(s,t;\sigma,\tau) \simeq \frac{1}{16}\Theta^{\text{flat}}_4(s,t;\sigma,\tau)\me(s,t;\sigma,\tau),~\text{ with }\Theta^{\text{flat}}_4(s,t;\sigma,\tau)=(tu+st\sigma+su\tau)^2.
\end{align}}
\begin{align}\label{eq:Penedones_formula}
	\lim_{L\rightarrow\infty}M(L^2s,L^2t)=\frac{L^{-1}}{\Gamma(\Sigma-2)}\int_0^\infty d\beta\beta^{\Sigma-3}e^{-\beta}\ca_{\text{flat}}\Big(\frac{2\beta s}{L^2},\frac{2\beta t}{L^2}\Big),
\end{align}
where $L$ is the radius of AdS, $\Sigma$ is half the sum of external charges, $\Sigma=\frac{p_1+p_2+p_3+p_4}{2}$, and in our particular case $\ca_{\text{flat}}$ is the ten-dimensional type IIB scattering amplitude of four super-gravitons in flat space.

Here we will follow the logic of~\cite{Chester:2018dga} and extend this formula to four-point functions with arbitrary Kaluza-Klein modes as external operators,\footnote{In the present case of AdS$_5\times$S$^5$, this was done for the $\corr22pp$ family of correlators in~\cite{Alday:2018pdi,Binder:2019jwn} and later generalised to the case of arbitrary external charges in~\cite{Drummond:2019odu}. Note that an interesting new type of flat space limit called the `large $p$ limit' has been recently proposed in~\cite{Aprile:2020luw}, where additionally to $s$ and $t$ one also takes the new Mellin variables corresponding to $\sigma$, $\tau$ as well as the external charges to be large.} repeating the analysis given already in~\cite{Drummond:2019odu}. Starting from the above ten-dimensional expression in flat space, we need to restrict the kinematics to the five-plane $\mathbb{R}^5\simeq\text{AdS}_5|_{L\rightarrow\infty}$ by integrating over the S$^5$ wavefunctions of the Kaluza-Klein modes dual to $\op$, where the integration over S$^5$ yields an additional factor of $L^5$. Denoting the ten-dimensional amplitude in transverse kinematics by $\ca^{(10)}_{\perp}(s,t;\sigma,\tau)$, the relation~\eqref{eq:Penedones_formula} can be inverted to give
\begin{align}\label{eq:flat_space_limit}
	\ca^{(10)}_{\perp}(s,t;\sigma,\tau) = \frac{\Theta^{\text{flat}}_4(s,t;\sigma,\tau)}{16~\mathcal{N}_{\ca}}~\me_{\lim}(s,t;\sigma,\tau),
\end{align}
where $\me_{\lim}$ implements the large $s,t$ limit and is given by
\begin{align}\label{eq:m_lim}
	\me_{\lim}(s,t;\sigma,\tau)=\Gamma(\Sigma-2)\lim_{L\rightarrow\infty}L^{14}\int_{-i\infty}^{+i\infty}\frac{d\alpha}{2\pi i}~\alpha^{-(\Sigma+2)}e^{\alpha}~\me\Big(\frac{L^2s}{2\alpha},\frac{L^2t}{2\alpha};\sigma,\tau\Big).
\end{align}
Note that we made use of equation~\eqref{eq:full_Mellin_amplitude} to replace $M(s,t)$ with the reduced Mellin amplitude $\me(s,t;\sigma,\tau)$. Furthermore, the normalisation factor $\mathcal{N}_{\ca}$ depends only on the sum of charges through $\Sigma$ and additionally has a non-trivial $(\sigma,\tau)$ dependence. In the conventions used here, it is given by
\begin{align}\label{eq:flat_space_normalisation}
	\mathcal{N}_{\ca} = -\frac{(\alpha')^3}{32\pi^5}\,\frac{B^{\text{sugra}}_{p_i}(\sigma,\tau)}{\Sigma-2}.
\end{align}
The dependence on the $su(4)$ cross-ratios is fully captured by the factor $B_{p_i}^{\text{sugra}}$, which follows from the large $s,t$ limit of the tree-level supergravity amplitude $\me^{(1,0)}$: 
\begin{align}\label{eq:B_sugra}
	B_{p_i}^{\text{sugra}}(\sigma,\tau) = \sum_{i,j\geq0}\frac{1}{i!j!k!}\frac{8p_1p_2p_3p_4}{\big(\frac{p_{43}+p_{21}}{2}+i\big)!\big(\frac{p_{43}-p_{21}}{2}+j\big)!\big(\frac{|p_1+p_2-p_3-p_4|}{2}+k\big)!}\sigma^i\tau^j,
\end{align}
with $p_{ij}=p_i-p_j$, $k=p_3+\min\left\{0,\frac{p_1+p_2-p_3-p_4}{2}\right\}-i-j-2$ and the range of $i,j$ is such that $k\geq0$ in the sum.

\subsection{Matching the genus-one string amplitude}
Let us finally demonstrate that in the flat space limit the constructed one-loop Mellin amplitudes $\me_p^{(2,3)}$ and $\me_{3333}^{(2,3)}$ match the ten-dimensional type IIB closed string theory scattering amplitude.

After performing the $\alpha$-integration and taking the limit $L\rightarrow\infty$ of equation~\eqref{eq:m_lim}, the two amplitudes read
\begin{align}
\begin{split}
	\me^{(2,3)}_{p,\lim}&=-L^{22}~\frac{p\,\zeta_3}{5760(p-2)!}\,\big(s^4\log(-s)+\text{crossed}\big),\\
	\me^{(2,3)}_{3333,\lim}&=-L^{22}~\frac{9\zeta_3}{10240}\,(1+\sigma+\tau)\,\big(s^4\log(-s)+\text{crossed}\big),
\end{split}
\end{align}
where we have used that in the limit $\widetilde{\psi}_0(x)\rightarrow\log(x)$. Next, we divide by the respective normalisation factors, where $\mathcal{N}_{\ca}$ depends on the sum of external charges and for the $\langle\ott\ott\ott\ott\rangle$ correlator has also a non-trivial dependence on the internal R-symmetry variables through the polynomial $B^{\text{sugra}}_{p_i}(\sigma,\tau)$ given in~\eqref{eq:B_sugra}. For the cases at hand, we have
\begin{align}
	B^{\text{sugra}}_{22pp} = \frac{32p^2}{(p-2)!},\qquad B^{\text{sugra}}_{3333}=648(1+\sigma+\tau).
\end{align}
As expected, we find that all dependence on the $su(4)$ cross-ratios $(\sigma,\tau)$ and the external charges cancels, such that both $\me_p^{(2,3)}$ and $\me_{3333}^{(2,3)}$ have the same flat space limit. After reinstating the factors of $a^2$ and $\ll$, the RHS of equation~\eqref{eq:flat_space_limit} is given up to an overall normalisation by
\begin{align}\label{eq:flat_space_result}
	a^2\ll~\frac{\Theta^{\text{flat}}_4(s,t;\sigma,\tau)}{16}~L^{22}\,\zeta_3\, \big(s^4\log(-s)+\text{crossed}\big).
\end{align}
Lastly, we need to convert the CFT quantities $a$ and $\lambda$ from the double into string theory quantities. According to the AdS/CFT dictionary, we have
\begin{align}\label{eq:dictionary}
	a=\frac{1}{N^2-1}~\sim~\frac{g_s^2\,\alpha'^4}{L^8},\qquad \l^{-\frac{1}{2}}~\sim~\frac{\alpha'}{L^2},
\end{align}
such that the factor of $L^{22}$ in~\eqref{eq:flat_space_result} is precisely cancelled.

The resulting expression should be compared to the type IIB flat space string amplitude in transverse kinematics, $\ca^{(10)}_{\perp}$. The four-graviton scattering amplitude admits the following genus expansion,
\begin{align}\label{eq:flat_space_amplitude}
	\ca^{\text{(10)}}_{\perp} &= \kappa_{10}^2\,g_s^4\,\Big(\frac{1}{g_s^2}\ca^{\text{tree}} + 2\pi \left(\ca^{\text{genus-1}}_{\text{an}}+\ca^{\text{genus-1}}_{\text{non-an}}\right)+O(g_s^2)\Big),
\end{align}
The low-energy expansion of the tree-level term $\ca^{\text{tree}}$ takes the form\footnote{Here, $\hat{\sigma}_2$ and $\hat{\sigma}_3$ are defined as
\begin{align}
	\hat{\sigma}_2=\Big(\frac{\alpha'}{4}\Big)^2\,(s^2+t^2+u^2),\qquad \hat{\sigma}_3=3\,\Big(\frac{\alpha'}{4}\Big)^3\,stu,
\end{align}
where $s$, $t$ and $u$ are the usual ten-dimensional Mandelstam invariants obeying $s+t+u=0$.}
\begin{align}\label{eq:tree-level_expansion}
	\ca^{\text{tree}}=\mathcal{R}^4\,\Big(\frac{3}{\hat{\sigma}_3}+2\zeta_3+\hat{\sigma}_2\zeta_5+\frac{2}{3}\hat{\sigma}_3(\zeta_3)^2+\ldots\Big),
\end{align}
while the genus-one terms are expanded as~\cite{Green:1999pv,Green:2008uj}
\begin{align}\label{eq:genus-one_analytic}
	\mathcal{A}^{\text{genus-1}}_{\text{an}} &= \frac{\pi}{3}\,\Big(1+0\hat{\sigma}_2+\frac{\zeta_3}{3}\hat{\sigma}_3+0\hat{\sigma}_2^2+\frac{97}{1080}\zeta_5\hat{\sigma}_2\hat{\sigma}_3+\ldots\Big)\,\mathcal{R}^4,\\
\begin{split}\label{eq:genus-one_non-analytic}
	\mathcal{A}^{\text{genus-1}}_{\text{non-an}} &= \mathcal{A}^{\text{genus-1}}_{\text{sugra}}+\Big(\frac{\alpha'}{4}\Big)^4\,\frac{4\zeta_3\pi}{45}\,\Big[s^4\log\Big(-\frac{\alpha' s}{\mu_4}\Big)+\text{crossing}\Big]\,\mathcal{R}^4\\
	&\quad+\Big(\frac{\alpha'}{4}\Big)^6\,\frac{\zeta_5\pi}{2520}\,\Big[(87s^6+s^4(t-u)^2)\log\Big(-\frac{\alpha' s}{\mu_6}\Big)+\text{crossing}\Big]\,\mathcal{R}^4+\ldots.
\end{split}
\end{align}
Note that all of the above terms share a common factor of $\mathcal{R}^4$, and it has been shown in~\cite{Chester:2018dga} that in transverse kinematics it is given by
\begin{align}
	\mathcal{R}^4_{\perp} = \frac{\Theta^{\text{flat}}_4(s,t;\sigma,\tau)}{16},
\end{align}
and therefore it cancels against the identical overall factor in equation~\eqref{eq:flat_space_limit}. With this in mind, we see that the flat space limit of $\me_p^{(2,3)}$ and $\me_{3333}^{(2,3)}$ given in equation~\eqref{eq:flat_space_result} exactly match the structure of the $(\alpha')^4$ term of the non-analytic genus-one expansion~\eqref{eq:genus-one_non-analytic}. This constitutes a non-trivial check on the Mellin amplitudes derived in this paper.

To conclude, let us also briefly comment on the next order in the $1/\lambda$ expansion. At order $\lll$, so far only the $\fourtwo$ one-loop correlator has been derived, see~\cite{Alday:2018kkw} and~\cite{Drummond:2019hel} for more details. Let us show here that its Mellin amplitude $\me_2^{(2,5)}$ correctly matches the corresponding term in the flat space string amplitude. After performing the $\alpha$-integration of equation~\eqref{eq:m_lim} and taking the large $L$ limit, we find
\begin{align}
	\me^{(2,5)}_{\lim}\sim a^2\lll\, L^{26}\,\zeta_5\big(s^4(22s^2+st+t^2)\log(-s)+\text{crossing}\big),
\end{align}
which non-trivially matches the $(\alpha')^6$ term of the genus-one low-energy expansion in the second line of~\eqref{eq:genus-one_non-analytic} upon using the identity $u=-s-t$.

\section*{Acknowledgements}
JMD and HP acknowledge support from ERC Consolidator grant 648630 IQFT. RG is supported by an STFC studentship.

\begin{appendix}
\section{Subleading OPE data in the window-region}\setcounter{equation}{0}\label{ap:OPE}
In this appendix we collect some new results for subleading three-point functions in the window-region. We were able to find a closed formula for all order $\ll$ string corrected three-point functions in the singlet channel, while the supergravity ones turn out to be of a more complicated form and hence we only present results for the first few twists.

\subsection{Results for $C^{(1,3)}$}
The string corrected three-point functions are non-vanishing only for degeneracy label $i=1$ and spin $\ell=0$ (mirroring the behaviour of the string anomalous dimensions $\eta^{(1,3)}_i,$ see~\eqref{eq:anom_dim_-3/2}). In the window-region, i.e. for twists $t<p$, they are given by
\begin{align}\label{eq:C31_window}
	C^{(1,3)}_{pp,t,1} =C^{(0,0)}_{22,t,1}~\frac{\zeta_3}{1680}~\frac{(-1)^t t^2 p (t-1)(1+t)^3(2+t)^2(3+t) \Gamma(p-t) \Gamma(p+2+t)}{\Gamma(p-1) \Gamma(p)}~\delta_{\ell,0}.
\end{align}

\subsection{Results for $C^{(1,0)}$}
We have calculated the subleading supergravity three-point functions up to twist sixteen and specialising to spin $\ell=0$ (note that, unlike the $C^{(1,3)}$, the $C^{(1,0)}$ have infinite spin support). In the window-region $t<p$, they take the form
\begin{align}
	C^{(1,0)}_{pp,t,i}|_{\ell=0} = C^{(0,0)}_{22,t,i}\times \frac{p^2(1+p)(2+p)}{(p-t)_{t-1}}\times\widetilde{C}^{(1,0)}_{pp,t,i},
\end{align}
where we have
\begin{align}
	\widetilde{C}^{(1,0)}_{pp,t=2} = \left\{\frac{1}{6}\right\},
\end{align}
\begin{align}
	\widetilde{C}^{(1,0)}_{pp,t=3} = \left\{\frac{1}{2} (-2 p-1),\frac{1}{56} (7 p-19)\right\},
\end{align}
\begin{align}
	\widetilde{C}^{(1,0)}_{pp,t=4} = \left\{\frac{7}{2} \left(p^2+p+4\right),\frac{1}{40} \left(-31 p^2+140 p-106\right),\frac{1}{30} \left(6 p^2-38 p+57\right)\right\},
\end{align}
\begin{align}
\begin{split}
	\widetilde{C}^{(1,0)}_{pp,t=5} = \bigg\{& -\frac{14}{3} (2 p+1) \left(p^2+p+15\right),\frac{1}{6} \left(19 p^3-129 p^2+263 p-267\right),\\
	& \frac{1}{450} \left(-603 p^3+5756 p^2-16623 p+13996\right),\frac{1}{132} \left(55 p^3-600 p^2+2081 p-2272\right) \bigg\} ,
\end{split}
\end{align}
\begin{align}
\begin{split}
	\widetilde{C}^{(1,0)}_{pp,t=6} = \bigg\{& 21 \left(p^2+p+2\right) \left(p^2+p+36\right),-\frac{3}{100}  \left(335 p^4-3200 p^3+12111 p^2-26150 p+15696\right),\\
	& \frac{1}{1750}\left(10889 p^4-145950 p^3+684667 p^2-1363950 p+1041408\right),\\
	&\frac{1}{1100}\left(-3355 p^4+51400 p^3-278551 p^2+626250 p-487416\right),\\
	& \frac{1}{143} \left(143 p^4-2354 p^3+13937 p^2-34926 p+30984\right) \bigg\},
\end{split}
\end{align}
\begin{align}
\begin{split}
	\widetilde{C}^{(1,0)}_{pp,t=7} = \bigg\{ & -21 (2 p+1) \left(p^4+2 p^3+79 p^2+78 p+504\right),\\
	& \frac{3}{100} \left(895 p^5-11465 p^4+68387 p^3-239479 p^2+348582 p-257544\right),\\
	& \frac{1}{500} \left(-11353 p^5+203426 p^4-1403999 p^3+4795558 p^2-8209656 p+5325552\right),\\
	& \frac{1}{1100}\left(17435 p^5-356845 p^4+2777467 p^3-10274939 p^2+18164322 p-12423024\right),\\
	& \frac{1}{143} \left(-1144 p^5+25157 p^4-211574 p^3+845563 p^2-1595682 p+1130760\right),\\
	& \frac{3}{1144} \left(1001 p^5-23023 p^4+204061 p^3-866657 p^2+1752498 p-1339560\right) \bigg\},
\end{split}
\end{align}
\begin{align}
\begin{split}
	 & \widetilde{C}^{(1,0)}_{pp,t=8}  = \bigg\{ 77 \left(p^6+3 p^5+145 p^4+285 p^3+2374 p^2+2232 p+2880\right),\\
	 & -\frac{11}{4}  \left(23 p^6-381 p^5+3263 p^4-16539 p^3+42146 p^2-67248 p+35136\right),\\
	 & \frac{11}{350} \left(2219 p^6-51247 p^5+487591 p^4-2490089 p^3+7129590 p^2-10524192 p+6399168\right),\\
	 & -\frac{1}{1764}(p-4) \left(114929 p^5-2569387 p^4+21708697 p^3-87056237 p^2+167481198 p-121386960\right),\\
	 & \frac{1}{1911}\left(87919 p^6-2488899 p^5+28178263 p^4-162892653 p^3+506300194 p^2-802071144 p+506306880\right),\\
	 & -\frac{3}{728} \left(5551 p^6-164346 p^5+1952327 p^4-11860562 p^3+38682426 p^2-63904876 p+41580720\right),\\
	 & \frac{11}{1326} \left(884 p^6-26988 p^5+331955 p^4-2097030 p^3+7141121 p^2-12358422 p+8435160\right)\bigg\}.
\end{split}
\end{align}
\end{appendix}

\end{document}